\documentclass[doublecol]{epl2}
\usepackage{amssymb}
\usepackage{mathrsfs}

\title{Towards a temporal network analysis of interactive WiFi users\footnote{The first two authors contributed equally to this paper.}} \shorttitle{Towards a temporal network analysis of interactive WiFi users} 

\author{Yan Zhang \and Lin Wang \and Yi-Qing Zhang \and Xiang Li\thanks{\email{lix@fudan.edu.cn}}}
\shortauthor{Y. Zhang \etal}

\institute{Adaptive Networks and Control Lab, Department
of Electronic Engineering, Fudan University, Shanghai 200433, P.R.China}

\abstract{Complex networks are used to depict topological features of complex systems. The structure of a network characterizes the
interactions among elements of the system, and facilitates the study of many dynamical processes taking place on it.
In previous investigations, the topological infrastructure underlying dynamical systems is simplified as a
static and invariable skeleton. However, this assumption cannot cover the temporal features of many time-evolution
networks, whose components are evolving and mutating. In this letter, utilizing the log data of WiFi
users in a Chinese university campus, we infuse the temporal dimension into the construction of dynamical human
contact network. By quantitative comparison with the traditional aggregation approach, we find that the temporal
contact network differs in many features, e.g., the reachability, the path length distribution. We conclude that the
correlation between temporal path length and duration is not only determined by their definitions, but also influenced by the
micro-dynamical features of human activities under certain social circumstance as well. The time order
of individuals' interaction events plays a critical role in understanding many dynamical processes via human close
proximity interactions studied in this letter. Besides, our study also provides a promising measure to identify
the potential superspreaders by distinguishing the nodes functioning as the relay hub.}

\begin{document}
\maketitle
\section{Introduction}{Complex network theory has been widely applied in many fields of science and
engineering\cite{RMP74,TCASMAG36}. Many complex systems are characterized by models of complex networks, where
nodes represent the cells of systems, and edges describe the interactions among them. Recent years, the study of
dynamical processes on complex network has received extensive attention\cite{NPV}, e.g. epidemic spread \cite{PRL863200,PCBe1001109,PRL218701,PLOSONE6e21197}, population dynamics\cite{EPL201248001,PRL028702}, and
Internet packet routing\cite{PRL218702,EPL14003}. In previous investigations, the topological infrastructure underlying
dynamical systems is generally simplified as a static and invariable skeleton,
where nodes and edges are assumed to be permanent entities (although the edge weights might change dynamically). This
simplification stems from the fact that detailed temporal information of structure evolution is untouchable
according to some technology deficiency, or the variation of network structure is not frequent enough to influence
the dynamical processes(e.g., the relation between the highway network and the commuting traffic). Obviously, this
assumption cannot cover the temporal features of many time-evolution networks, whose components are evolving
and mutating\cite{PR,arxiv2,PRE016105,PA3881007,PRE025102}}. Taking human contact networks\cite{PNAS10722020,PLOSONE5e11596,JTB271,plosonezyq} for
example, people do not keep on interacting with their neighbors perpetually, and the time ordering is intrinsically
embedded in the human close proximity interactions(CPIs)(in other words, each individual does not simultaneously contact all
the neighbors, the times when the interactions are active determine the temporal sequence). Traditionally, a network is
regarded as the underlying infrastructure of dynamical processes. Introducing the temporal dimension may help us infuse
the temporal information about dynamical events into the network construction\cite{PR,arxiv2,PRE016105,PA3881007,PRE025102}.

Recently, the analysis of temporal features of complex communication networks, especially the human contact networks, have
received a boost with the advance of information technology, e.g., wireless communication. Newly created digital instruments not only reshape our daily life, but also record tremendous digital data produced by human daily activities, which can be utilized to analyze human behaviors
\cite{PNAS10722020,PLOSONE5e11596,JTB271,plosonezyq}. WiFi, as a ubiquitous wireless accessing technology, has been widely deployed in human
daily circumstances: Actually, the ``WiFi-Free'' sign can be found in nearly every corner of urban areas, and the notion of ``WiFi-City'' becomes reality. The commercial WiFi system provides us a powerful tool to collect digital traces of a huge
population without artificial interference. In this letter, with the digital traces automatically recorded by
the WiFi control system of a Chinese university campus, we construct a contact network of human CPIs  from
the temporal perspective, which mainly embraces the concept of vector clocks widely used in distributed
computing\cite{ACM558}. By quantitative comparison with the traditional aggregation networks, we find that the temporal
contact network differs in many features, e.g., the reachability, the path length distributions. We conclude that the correlation between temporal path length and duration is not only determined by their definitions, but also influenced by the micro-dynamical features of human activities under certain social circumstance(we use two more datasets to check this result). The time order of interaction events plays an important role in understanding many dynamical processes via human CPIs, e.g., epidemic spread, information diffusion. Besides, our study provides a promising measure to identify the potential superspreaders by distinguishing the nodes functioning as the relay hub.

\section{\textbf{Background and data description}}
The WiFi system in this study is deployed at the Handan campus of Fudan University in Shanghai, China, where all
administrative, athletic, academic and teaching buildings have been covered by wireless access points (WAPs).
This system automatically records users' access-related events, which are marked by the information pertaining to the
Media Access Control (MAC) address of users' mobile devices, the online and offline time, and the MAC address of WAPs.
We mainly select WiFi data recorded from all the 6 teaching buildings for two reasons: (i) the 6 teaching buildings are open to
all the campus members because almost all curricula are scheduled here, while other buildings are limited or partially limited
 to certain campus members; (ii) in the 6 public teaching buildings, WiFi users
seldom leave their mobile devices for a long period, whereas it is regular that a machine is still connected
to the WAP though the user has already left the ``secured" office. Our study uses the data from 14th October to 25th
November, covering about 14000 WiFi users.

In order to extract the information of CPIs from the data, an assumption is put forward at first: any WiFi
device seeing the same AP infers a close indoor proximity interaction among the devices owners \cite{ITMC1536}. The intrinsic reason of this assumption
 lies in the fact that in a close indoor circumstance people have a high probability to directly communicate
with each other or have some relationship via indirect interactions. For instance, a respiratory disease, e.g., the SARS (
the survival period of the SARS virus is about 24 hours, which is long enough to infect people in a close indoor
circumstance) might propagate from person to person; a computer virus might bypass the
communication protocols and spread from one device to others through the WAP or other peer to peer manners such as Bluetooth\cite{PNAS1061318}.

We define ${e_i}=({V_i},{t_{i1}},{t_{i2}})$ as an interaction event(IE), where ${V_i}$ denotes the set of users
connecting to the same WAP simultaneously, and any user $v\in {V_i}$ interacts with others in ${V_i}$  during the
time period from ${t_{i1}}$ to ${t_{i2}}$. According to this treatment, we translate the WiFi access-related logs
into human interaction events.

\begin{figure}[h!]
\begin{center}
\includegraphics[width=6cm]{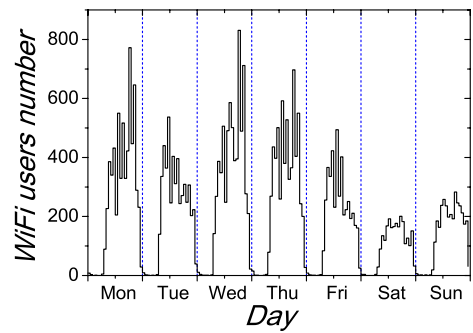}
\caption{ Number of WiFi users in each hour in the week(19th October to 25th October).}\label{fig.1}
\end{center}
\end{figure}

We first partition the whole dataset into several subsets, each of which contains human activities at each week in the observed
month. Counting the number of WiFi users at each hour of any subset, we find that there are circadian patterns
 underlying the contact activities. As shown in Fig.\ref{fig.1}, the average amount of WiFi users from Monday to Friday is evidently more than that in the weekend. This phenomenon indicates that users' WiFi accessing behaviors are in accordance with the university weekly curriculum
schedules. Strikingly, we also observe that the amount of WiFi users intensely fluctuates as time elapses on each workday, while this fluctuation has been largely weakened at weekends. During the lunch and dinner time on a workday,
most people go to the dining hall, thus few still keep online. Meanwhile, students and teachers always need to change
classrooms at the break time according to the workday curriculum schedules. In the night of workdays, students commonly
stay in classrooms with their laptops to do their homework or for entertainment. Since most of their mobile devices get connected
to the WAPs, the users' number peaks at night (Tuesday and Friday are exceptional as there is no class arranged at
Tuesday afternoon, and many native students are homeward at Friday night). Therefore, the fluctuation of the number of WiFi users
indicates that the natural rhythm of individuals' daily activities is in accordance with the university daily schedules. The teaching buildings
are open from 7AM to 11PM, and people have to leave these building before 11PM. Therefore, we only take into account the IEs occurring in the 16 opening hours every day hereafter.

\begin{figure}[h!]
\begin{center}
\includegraphics[width=7cm]{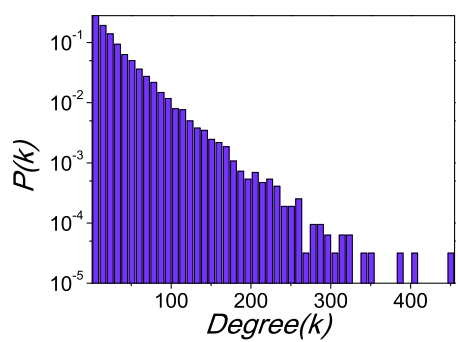}
\caption{ The degree distribution of the ACN with a time window at the weekly level. The distribution is averaged over all weeks..}\label{fig.2}
\end{center}
\end{figure}

From the traditional aggregated network perspective, each WiFi user is denoted as a node, and an edge links any two nodes
if there is at least one IE that involves them both taking place under the observation epoch. Each edge is weighted by the number of
contacts (or the total contact duration). The strength of a node is the sum of the weight of edges departing from the node. Aggregating all the IEs in each week, we can build the aggregated contact network(ACN) at the weekly level. Fig.\ref{fig.2} presents the node degree distribution of weekly aggregated networks averaged over all weeks. The distribution is exponentially distributed, indicating that most users only have a limited number of contacts, which has also been reported in \cite{JTB271}. We alter the time window of the aggregation process(e.g.,  one day, one month), and find that the exponential trend is robust.

\section {\textbf{Constructing the temporal contact network}}
\begin{figure}[h!]
\begin{center}
\includegraphics[width=7cm]{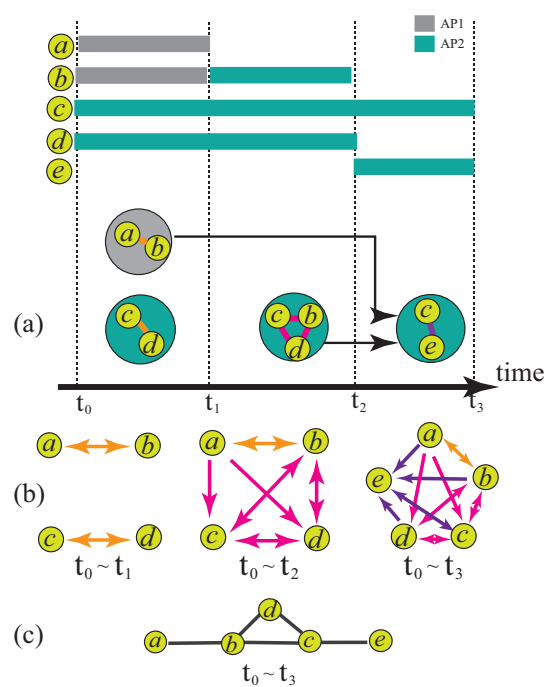}
\caption{(color online)Schematic illustration of the TCN and ACN construction.
(a) WiFi accessing logs are translated into the corresponding IEs. The black arrowed lines show the temporal contacts between individuals in different IEs.
(b) The construction of TCN. The edges are colored according to the latest time that they are updated.
(c) The construction of ACN at the observed duration $t_{0}\thicksim t_{3}$
}\label{fig.3}
\end{center}
\end{figure}

From the temporal perspective, we first define a temporal contact(TC): user $i$ makes a TC with $j$ if there is a
sequence of IEs with non-decreasing time, linking $i$ and $j$ \cite{JCSS76036113}. Because time creates its
own dimension in the events sequences, the TCs are directed. For instance, in Fig.\ref{fig.3}, user $a$ interacts with $b$
before $b$ interacts with $c$. There is only a TC from $a$ to $c$, and $c$ cannot make a TC with $a$,
due to the fact that nothing(e.g., information) could be propagated from $c$ to $a$ via $b$. In a temporal contact network(TCN), an edge links any two nodes if they have at least one TC.

Before a given observation time $t$, there might exist many TCs linking users $i, j$. We define $\phi_{i,j}(t)$ as the time of the inception of the latest TC from $i$ to $j$ before $t$. Considering that a new TC from $i$ to $j$ is created at time $t'$, we define the temporal path duration as $\tau_{i,j}(t')=t'-\phi_{i,j}(t')$, which measures the time interval consumed by the corresponding temporal path. Before the given observation time $t$, the temporal path length $\theta_{i,j}(t)$ is defined as the shortest length of all the latest TCs(the least number of IEs of all the latest TCs). Therefore, the edges in the TCN can be weighted by the temporal path length or duration(The detailed algorithms of calculating $\phi_{i,j}(t)$ and $\theta_{i,j}(t)$ are given in Appendix).

\section {\textbf{Analysis}}\subsection{Reachability and Path Lengths}

In the ACN, the size $N_{i}$ of the component that any given node $i$ belongs to indicates an upper bound of the number of nodes that can be influenced by the spreading dynamics(e.g, virus transmission) launched from $i$. $N_{i}$ characterizes the reachability of node $i$ in the aggregated network. Employing the casual timing of the IEs to construct the temporal version
of the contact network, we first measure network reachability. With any given node $i$, we denote the number of nodes that can be temporally contacted by $i$ within a given observation period $\Delta t=t_{2}-t_{1}$ as the reachability of $i$\cite{PRE046119}, which is equal to the number
of elements in the set $\{ {\phi _{i,j}}(t_2)\mid {\phi _{i,j}}(t_2) > {t_1} \ j \in V\}$. The average(maximum) network reachability is denoted by the mean(maximum) value of all nodes' reachability.

\begin{figure}[h!]
\begin{center}
\includegraphics[width=8cm]{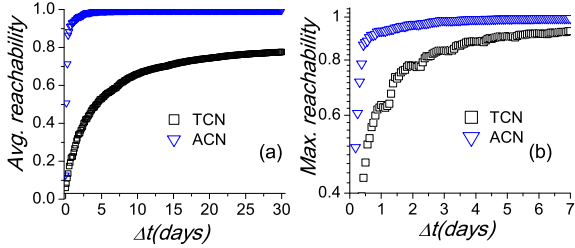}
\caption{(a)Comparison of the average reachability between the ACN and TCN. (b)Comparison of the maximal reachability between the ACN and TCN.}\label{fig.4}
\end{center}
\end{figure}

Fig.\ref{fig.4} compares the network reachability(normalized by the network size) between the TCN and ACN with the same observation period. Both the average and maximum value show that the ACN's reachability is larger, especially as $\Delta t\rightarrow 0$. The reachability of ACN quickly attains its saturation as $\Delta t$ increases, while the saturation of the reachability of TCN is much slower. The saturated value of the average reachability of the TCN is much smaller than that of the ACN. Hence the temporal dimension provides a tighter upper bound for the network reachability.

We also compare the distribution of the shortest path lengths of the ACN with that of the TCN. In the ACN, the shortest path length $d^{a}$ denotes the topological distance between any given source and destination, while, in the TCN, the temporal version of the shortest path length $d^{t}$ denotes the least number of the IEs of a given fastest TC.  We find that the distribution of $d^{t}$ is broader than that of $d^{a}$, and the mean value of $d^{t}$ is larger. Besides, the maximum of $d^{t}$ is about two times larger than that of $d^{a}$.

The difference in reachability and path length manifests the discrepancy of the structure between ACN and TCN. In the following, we mainly study the features of TCN.

\begin{figure}[h!]
\begin{center}
\includegraphics[width=7cm]{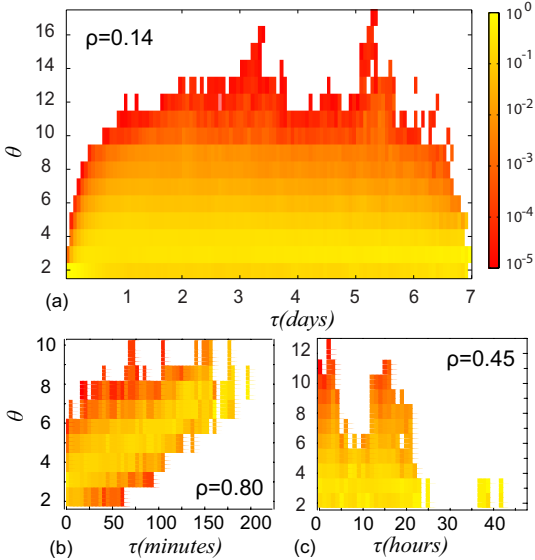}
\caption{(color online)The correlation of the temporal path duration $\tau$ and duration $\theta$ for (a) the WiFi system under study, (b) the SG exhibition, (c) the HT09 conference. In panel Fig.\ref{fig.5}(a), the conditional probability $p(\theta|\tau)$ is averaged at the weekly level.}\label{fig.5}
\end{center}
\end{figure}

\subsection{The correlation between path's length and duration}

In the ACN, the time spent on each edge can be regarded as identical, thus it is evident that the longer the path length is, the more time
it takes to build the path. In the TCN, the time consumed by the fastest TCs from person to person is characterized by the temporal path duration $\tau$, and the least number of IEs involved in the TCs is denoted by the temporal path length $\theta$. Whether or not the positive correlation between path length and duration in the ACN can also be found in the TCN remains unclear. Fig.\ref{fig.5}(a) presents the conditional probability distribution $p(\theta|\tau)$ of the WiFi deployment, which does not shows a positive correlation, but reveals the fact that considerable fluctuations of $\theta$ exist with any given $\tau$.  For instance, when we fix TCs' temporal path duration to 1 day, their temporal path lengths vary dramatically in $10^{4}$ orders. The Spearman correlation coefficient $\rho$ between $\theta$ and $\tau$ is 0.14, indicating that the strong positive correlation between path length and duration is absent in the TCN.

 Particularly, the definition of temporal path length and duration require us to select the fastest TCs with different inception time before measuring the length or duration. If there are several TCs pointing to a given destination node, one path duration may different from others though their path length is equivalent. It is also possible that in a given time period there exist more than one path between a given source and destination(their temporal path duration is equivalent) due to the fact that distinct sequences of IEs link them(individuals may present in many IEs in the given time period)\cite{plosonezyq}.

Furthermore, the correlation between temporal path length and duration is not only determined by their definitions, but is also influenced by the micro-dynamical features of human activities under certain social circumstance. To examine the impact of social circumstances to the correlation feature, we utilize the longitudinal data on human face-to-face proximity contact in conference(HT09) and museum(SG exhibition)\cite{JTB271} collected by active Radio Frequency Identification Devices(RFID). The datasets are provided by the SocioPatterns Project(http://www.sociopatterns.org/, more details see Appendix).
The attendees in the HT09 conference shape a closed population like the university members in our study, while the visitors in the SG exhibition seldom repeat their visit day after day. The absence of strong correlation between the temporal path length and duration in HT09 conference(Fig.\ref{fig.5}(c), $\rho=0.45$) is similar to that of Fig.\ref{fig.5}(a), whereas the temporal path length obviously increases with the growth of the temporal path duration in the SG exhibition(Fig.\ref{fig.5}(b), $\rho=0.80$). The distinct correlation results from the fact that the social circumstance between HT09 and SG is different: In the SG exhibition, visitors typically spend a limit time period on each site, and touch different locations following a rather pre-defined route; in the HT09 conference, most attendees stay on-site during the entire program (a few days), move at will among limited areas such as conference hall, corridor for coffer breaks and so on. Our university members and the HT09 conference attendees shape closed populations. Individuals can encounter with each other more frequently than the scenario of an `open' system like the SG museum exhibition. Moreover, in a closed population, individuals seldom quit the system, thus they can contact with each other iteratively, whereas the visitors in the SG exhibition form an `open' circumstance with a flux of individuals streaming through the sighting, and they seldom return once leave. Therefore, the temporal path length elongates with the growth of path duration in the SG exhibition.

\subsection{The correlation of the temporal out-degrees and in-degrees}

\begin{figure}[h!]
\begin{center}
\includegraphics[width=7.5cm]{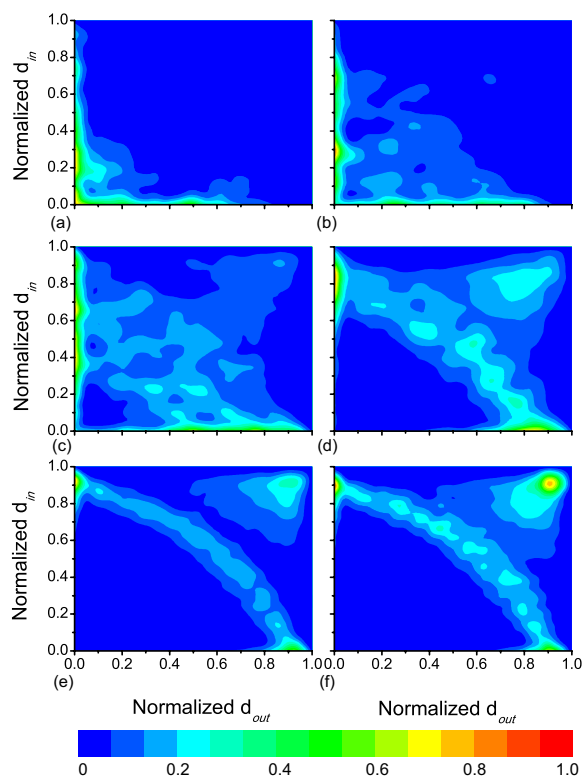}
\caption{(color online)The joint probability distribution of $d_{out}$ and $d_{in}$. When $\Delta t$ is larger that $7$ days, a set of data points always presents at the right-up corner. $\Delta t$ is
(a) $1$ day, (b) $2$ days, (c) $3$ days, (d) $5$ days, (e) $7$ days, (f) $8$ days.}\label{fig.6}
\end{center}
\end{figure}

In the TCN, the out-degree $d_{out,i}$ of any node $i$ quantifies the number of receptors temporally
affected by $i$, while the in-degree $d_{in,i}$ specifies the number of its potential inciters. The
correlation of $d_{out}$ and $d_{in}$ can be employed to distinguish the role of nodes in a dynamical
system \cite{JUH446,PRE046119}. Fig.\ref{fig.6}(a)-(f) present the joint probability distribution
$C^{\Delta t}(d_{out,i},d_{in,i})$ with different observation periods $\Delta t$. In each subgraph,
we uniformly partition the whole dataset into subsets with different $\Delta t$, measure the joint
probability distribution of each subset, and average them to calculate $C^{\Delta t}(d_{out,i},d_{in,i})$.

With small $\Delta t$ ($\Delta t=1$ day in Fig.\ref{fig.6}(a)), almost all data points reside in the lower triangular matrix, and many data
points are even sticking on the axes. It can be explained by the fact that in a short observation period, e.g., 1 day, only a
limited number of users keep frequently online all the time, while most individuals act according to their curriculum schedules, which
seldom cover the whole day. If some users access the WAPs in the morning, they can temporally
reach others logging online later in that day according to the above assumption and definitions, and they
are free from the temporal contacts of other subsequent users; if some people use WiFi at night, they can be temporally
influenced by other day time users(recall the fluctuations we observed in Fig.1). Occasionally, there are some individuals using WiFi both in the morning and at night,
but they do not inevitably become nodes with high in-degree and out-degree, because the value of degree
also depends on the number of nodes they directly interact with in the observation period. Therefore, a very
few individuals have a high potential to become nodes with large in- and out-degrees in a short
observation duration. As $\Delta t$ increases, the times that each user appears in the dataset also grow. The
forward and backward temporal influence of each user increases as time proceeds, and the limit of finite
observation period is gradually crippled. There are more and more data points emerging in the upper triangular
matrix of Fig.\ref{fig.6}(b)-(d)($\Delta t:$ 2 days$\rightarrow$ 6 days).When $\Delta t\geqslant$ 8 days, the
distribution of data points constantly generates two evidently nontrivial clusters: one contains the data
scattering along the diagonal, and the other contains the data anchoring in the upper right corner. They are mainly composed
of the frequently accessing users. There are also two obvious clusters of data points sticking on the axes, which contain the users
having less frequent connections. Though possessing large in-degree or out-degree, they are actually trivial hubs because
they lack the efficiency of transferring information or viruses.
In essence, the individuals presenting in the two nontrivial clusters, particularly, those in the upper right
corner in Fig.\ref{fig.6}, can play the role as relay hubs that are critical to spreading processes. Actually, when $\Delta t$
is small, there still exist a small amount of relay hubs(for example, see those in the right corner of Fig.\ref{fig.6}(b)),
and the similar pattern of the joint probability distributions with $\Delta t$ beyond
7 days might result from the weekly cycle of the curriculum schedules.

To examine whether or not each individual's role is invariable in the ACN and TCN, we rank individuals' degree in these two versions of contact networks, respectively. We calculate the Kendall's tau coefficient $\tau_{k}$ of the two rank series. Taking $\Delta t=7$ days as an example, the Kendall's tau coefficient between the ranking of in-degree(out-degree) in TCN and the ranking of degree in ACN is just $0.33$($0.30$) on average. It indicates that individuals' role may be different between the ACN and TCN. Nodes with high degrees in the ACN may be nodes with only hign in-degrees or out-degrees,  and many `leaf' individuals in the ACN are actually relay hubs in the TCN, whose role is often underestimated in the ACN analysis\cite{PLOSONE5e11596,plosonezyq}. The nodes that play the role as the relay hub in the TCN bear high force of infection, therefore they are more probable to act as superspreaders. The identification of influential spreaders is essential in designing optimal containment
strategies\cite{NP2010888}, and here we provide a direct method to identify potential superspreaders.

\section{\textbf{Conclusions}}

In this letter, we take advantage of the human digital traces automatically collected by the WiFi control system in a Chinese university campus. We map users' CPIs into TCNs. We analyze the infrastructure of the constructed dynamical system characterizing the CPIs among WiFi users from the perspective of temporal networks. This treatment infuses the temporal ordering of the WiFi users' interaction events into the network construction. By quantitative comparison with the aggregated contact networks, we uncover that the temporal contact network differs in many features, e.g., the reachability and the path length distribution. We find that the correlation between temporal path length and duration is not only determined by their definitions, but also influenced by the micro-dynamical features of human activities under social circumstances. Besides, we also provide a direct method to identify potential superspreaders, which does deserve further study in detail in future works, and may help us design more efficient containment plans.

\acknowledgments
We thank the SocioPatterns Project for sharing their high resolution data on human face-to-face proximity contact. This study is supported by
National Key Basic Research and Development Program (No.2010CB731403), the NCET program (No.NCET-09-0317) of China.

\section{\textbf{Appendix}}
\subsection{Calculation of $\phi(t)$ and $\theta(t)$}
To calculate $\phi(t)$ and $\theta(t)$, we modify the vector clocks based event-driven algorithm \cite{ACM558} by passing through all IEs ordered
by time. Take an IE ${e_i} = ({V_i},{t_{i1}},{t_{i2}})$ as an example, we specify the main steps of the algorithm as follows:

\noindent{\bf Step 1:}
Initialize the information about the nodes participating in the current IE.

\ \ \ \ \ \ \ \ \ \ \ $\phi_{u,u}(t)$=$t_{i2}$, $\theta_{u,u}(t)$=$0$, $\forall$ $u \in {V_i}$.

\noindent{\bf Step 2:}
Compare the vector clocks of the nodes in the IE, to find the latest temporal information.

\ \ \ \ \ \ \ \ \ \ \ ${\rm{T}}(v)=max{\{{\phi_{u,v}}(t) \mid\forall u\in {V_i}\}}$

\noindent{\bf Step 3:}
Find the shortest path length of the updating TCs: $L(v)=min\{ {\theta _{u,v}}(t) \mid\forall u\in{V_i}\ and\ {\phi_{u,v}}(t)={\rm{T}}(v)\}$.

\noindent{\bf Step 4:}
Update the temporal path information as

$\forall u\in {V_i}, \theta_{u,v}(t)=L(v)+1$, if ${\phi_{u,v}}(t) \ne {\rm{T}}(v)$ or ${\theta _{u,v}}(t) \ne L(v)$.
\ \ \ \  $\forall u\in {V_i}, \phi_{u,v}(t)={\rm{T}}(v)$.\newline
The update of $\phi(t)$ indicates the creating of a new TC, and the update of the corresponding element in $\Theta(t)$
records the shortest path length of the new TC.

\noindent{\bf Step 5:} Process the next IE and return to {\bf Step 1}.

Because each IE maintains a certain duration time, one IE often takes place before the
end of other IEs. So we split users' WiFi logs into shorter ones, to keep the time sequence of all concurrent IEs.

\subsection{The SocioPatterns project}
`SocioPatterns' is an interdisciplinary research collaboration that uses wireless technology to gather longitudinal data on human face-to-face proximity. It uses the data-driven approach to uncover fundamental patterns in social dynamics and human activity. The datasets we use are: The daily contact logs collected during the Infectious SocioPatterns event at the Science Gallery(SG) in Dublin; human face-to-face proximity data of about 110 conference attendees during the ACM Hypertext 2009(HT09) conference.



\end{document}